\documentstyle[aps,prd,preprint,floats,amsfonts]{revtex} \tighten
\begin{document}
%%%%%%%%%%%%%%%%%%%%%%%%%%%%%%%%%%%%%%%%%%%%%%%%%%%%%%%%%%%%%
\title{Photon Production from a Quark--Gluon Plasma}
\author{E.Quack\thanks{Present address: Theory Group, SLAC, Stanford
University, P.O. Box 4349, Stanford, CA 94309, USA; email:
quack@slac.stanford.edu} and P.A.Henning\thanks{e-mail:
P.Henning@gsi.de }} \address{Theory Group, Gesellschaft f\"ur
Schwerionenforschung (GSI) \\
        P.O.Box 110552, D-64220 Darmstadt, Germany}
\date{January 5, 1996, GSI-Preprint 95-42 (revised), hep-ph 9507273}
\maketitle
\begin{abstract}
In-medium interactions of a particle in a hot plasma are considered in 
the framework of thermal field theory. The 
formalism to calculate gauge invariant rates for photon and dilepton 
production from the medium is given. In the application to a QED plasma, 
astrophysical consequences are pointed out. The photon production rate from 
strongly interacting quarks in the quark--gluon plasma, which might be formed 
in ultrarelativistic heavy ion collisions, is calculated in the previously 
unaccessible regime of photon energies of the order of the plasma temperature.
 For temperatures below the chiral phase transition, an effective field 
theory incorporating dynamical chiral symmetry breaking is employed, and 
perturbative QCD at higher temperatures. A smooth transition between both 
regions is obtained. The relevance to the soft photon problem and to 
high energy heavy ion experiments is discussed. 
\end{abstract}
%%%%%%%%%%%%%%%%%%%%%%%%%%%%%%%%%%%%%%%%%%%%%%%%%%%%%%%%%%%%%%%%%%%%%%
\section{Introduction}
Considerable effort is invested in present and future experiments of 
ultrarelativistic heavy ion collisions (URHIC) in order to observe an 
excursion of the bulk of strongly interacting matter from the state 
of hadrons before the collision into the phase of a quark--gluon plasma (QGP) 
\cite{qm}. In order to see this shortlived state directly, one would 
like to observe photons emitted from the hot plasma, as well as dileptons. 
Since these probes interact only electromagnetically, 
their signal is not distorted by later interactions as 
are other particles which are studied for the same purpose. 

The experimental capability of measuring electromagnetic probes was 
demonstrated in the photon channel by Helios
\cite{helios}, WA80/98 \cite{wa80} and CERES \cite{cergam} as well as in the 
dilepton channel by Helios \cite{helios2} and CERES \cite{ceres}. 

The signal originating from the plasma phase is, however, buried under a 
background of photons from different origin such as from the decays of 
$\pi^0$ or $\eta$ or from hadronic reactions at a temperature 
comparable to that of the deconfined phase \cite{kap1}. After subtraction of 
these sources, a remaining signal seems to persist in part of the 
experimental analysis. At present, it is vividly discussed 
to what extent one can account for these data within more \cite{phot} or 
less \cite{eta} conventional physical pictures. 

However, our theoretical knowledge of the spectrum of electromagnetic probes 
from both the plasma as well as the hadronic phase is still uncertain to some 
extent. A better handle on these spectra from theoretical calculations is 
necessary in order to disentangle the various 
sources and to identify the phases reached during the collision. 
In particular for the soft part of electromagnetic radiation, 
this problem represents a challenge to theory in itself, due to the
nonperturbative nature of the photon emission process: Multiple
rescattering of the emitting particles and the Landau-Pomeranchuk-Migdal 
(LPM) effect play an important role in the medium for photon 
energies $E_{\gamma} \leq T$ \cite{wel94,kv95}, as well as for dileptons of 
an invariant mass in this range. 

This problem motivated the present work, in which we will investigate the 
production of photons and dileptons from a strongly interacting plasma at 
finite temperature. After a short sketch of the insufficiencies of existing 
calculations, we show how one can reach an 
improvement by taking thermal scattering and thereby spectral broadening 
of the emitting particles in the heat bath into account.
The problem is addressed in the framework of thermal field theory,
results are given for a QED plasma as well as for a QGP within a 
model incorporating dynamical chiral symmetry breaking. A part of the 
results has been presented already in a short paper \cite{pe}. 

In an $\alpha_s$ expansion, the lowest order of photon production proceeds 
via annihilation ($q\bar{q} \to g\gamma$) and Compton ($qg \to q\gamma$) 
processes. In next to leading order 
(NLO), numerous corrections to these processes arise, a complete calculation 
of the order ${\cal O}(\alpha\alpha_s^2)$ has been achieved in \cite{aur}. 
With an initial quark and gluon distribution specified by distribution 
functions $f_1(E_1)$ and $f_2(E_2)$, and the final state quark or gluons 
distribution $f_3(E_3)$, the production rate reads 
\begin{eqnarray}\nonumber
  R^0 &=& 
  E\frac{dN^0_{\gamma}}{d^3\bbox{p}}\\
\label{eqragen}
& =& {\cal N} \int \sum_{i=1}^4 
  \frac{d^3\bbox{p}_i}{2E_i(2\pi)^3} f_1(E_1) f_2(E_2) (2\pi)^4 
\delta^4(p_1+p_2+p_3-p_4) |M|^2 [1\mp f_3(E_3)] \; , 
\end{eqnarray}
where the last factor takes into account Pauli blocking or Bose enhancement 
of the quark or gluon in the final state, and $M$ stands for the elementary 
cross section considered. 

The production of hard (high $p_{\bot}$) photons in reactions of colliding 
hadrons has been calculated using `cold' parton distributions delivered by 
structure functions and using $M$ in NLO. For sufficiently high $p_{\bot}$, 
very good agreement with the corresponding data is reached \cite{cq}, 
only towards low $p_{\bot}$ some discrepancy has been reported \cite{hust}. 
This may hint at the insufficiency of using even NLO calculations in the soft 
regime, but may equally well be due to our still insufficient knowledge of 
the parton distribution functions in the relevant $x$ and $Q^2$ range, see 
\cite{eda} for an analysis. Even with this minor uncertainty, 
one has reached a very good quantitative understanding of photon production. 

Now let us look at the same processes in a plasma, where the partons have 
reached a thermal distribution. We will consider situations in which the 
spatial extension of the plasma is lower than the mean free path of the 
photons emitted, i.e.~we consider the emission of `white' radiation in 
contrast to thermal black body 
radiation. Due to the small size of nuclei compared to the mean free path of 
an electromagnetically interacting particle, this is always the case for 
heavy ion collisions. 
Using $M$ in lowest order, and taking thermal quark ($q$), $\bar{q}$ and 
gluon distributions of temperature $T$ results in a 
production rate $R$ (per unit volume element) as \cite{hwa} 
\begin{equation}
\label{eqralo}
  R^0 = 
  E\frac{dN^0_{\gamma}}{d^3\bbox{p}} = \frac{5}{9} 
  \,\frac{\alpha \alpha_s}{2 \pi^2}\, T^2\, {\mathrm e}^{-E/T}\,
   \left[ \log{\frac{E T}{m^2}} + c^0 \right]
\end{equation}
with some constant $c^0$. This rate diverges when $m \to 0$, 
which is the crucial limit of chiral symmetry restoration for 
strongly interacting quarks approaching the phase transition temperature. 
This unphysical divergence will eventually be shielded by medium effects 
on the emission process. 

A step towards the calculation of such medium effects has been the 
application of the Braaten--Pi\-sarski method of hard thermal loops 
\cite{brpi} to this problem \cite{kap1,bai1}. The resulting photon production 
rate is 
\begin{equation}
\label{eqbrpi}
  R^{\mbox{\footnotesize BP}} = 
  E\frac{dN^0_{\gamma}}{d^3\bbox{p}} = \frac{5}{9} 
  \,\frac{\alpha \alpha_s}{2 \pi^2}\, T^2\, {\mathrm e}^{-E/T}\,
   \log{\frac{c_1 E}{g^2 T}} 
\end{equation}
with a constant $c_1\sim 3$ and the strong coupling constant $g$. 
For $g^2 \sim c_1$, the term $\log(E/T)$ reminds us of the validity of this 
approach only in the region of $E_{\gamma} \gg T$.

A more detailed investigation of this infrared problem within the hard 
thermal loop (HTL) method of Braaten and Pisarski has been presented in 
\cite{BPS94,petit}. In these works, the production of soft photons was 
studied thoroughly. 
The origin of the infrared problem could be traced back to divergences 
which occur when the real photon is emitted collinear to a thermal gluon. 
For this reason, it was concluded in \cite{BPS94,petit} that no finite 
value for the production 
rate of soft photons can be obtained within the HTL method. 

This is the motivation for the present work. Physically, the HTL method 
takes into account the thermal masses particles acquire in the medium. 
In addition to that, we now also consider the thermal scattering of the 
partons, which  results in an energy uncertainty as the quark propagates. 
As we will show below, this is the dominant physical process to be considered 
for quarks emitting soft photons with energies $E_{\gamma} \leq T$. The 
thermal scattering 
leads to a finite lifetime (or nonzero spectral width) of every excitation
in the medium, described in analogy to the decay width of an excited state
\cite{NRT83}. 
One effect of such a spectral width is that it naturally removes the infrared 
divergences mentioned before, therefore enables us to calculate production 
rates for soft photons. Secondly, the energy ``uncertainty'', which
is actually a kind of Brownian motion, is directly related to 
the emission rate of thermal photons.

This paper is organized as follows. In the next section, we present the 
general formalism of thermally scattered particles in a heat bath. This 
includes the calculation of thermal widths as well as the photon production 
rate in a gauge invariant manner. Section II.C 
gives the comprehensive example of a fermion (quark) in a QED plasma. 
Although not being realistic for the QCD case, it allows for simpler and 
often analytic solutions and thus for a clear illustration of the relevant 
physics. 

We then turn to the case of the QCD plasma, section III. A crucial aspect 
of QCD at low temperatures is the breaking of chiral symmetry. 
Hence up to the chiral phase transition temperature we describe the plasma 
by the Nambu--Jona-Lasinio model, which incorporates this feature 
dynamically (Sect. III.A). In the subsequent part of this work, III B, 
we turn to high-temperature perturbative QCD, with 
temperature dependent strong coupling constant $\alpha_s$. 
Section III.C  gives the results for the photon production rates over the 
entire range of temperatures and of photon energies, and we discuss the 
relevance to a variety of experimental situations in section III.D. 
%%%%%%%%%%%%%%%%%%%%%%%%%%%%%%%%%%%%%%%%%%%%%%%%%%%%%%%%%%%%%%%%%%%%%
%
%  Section II
%
%%%%%%%%%%%%%%%%%%%%%%%%%%%%%%%%%%%%%%%%%%%%%%%%%%%%%%%%%%%%%%%%%%%%%
\section{Photon Radiation in Thermal Field Theory}
In this section, we first briefly recall the formalism of thermal field theory 
using spectral functions, outline how the self energy of a thermal particle 
is obtained in general, how it is related to the thermal width and how gauge 
invariant rates for photon production are obtained therefrom. We finally 
illustrate the achievements with the example of a fermion in a QED plasma. 
%%%%%%%%%%%%%%%%%%%%%%%%%%%%%%%%%%%%%%%%%%%%%%%%%%%%%%%%%%%%%%%%%%%%%
%  Subsection II a
%%%%%%%%%%%%%%%%%%%%%%%%%%%%%%%%%%%%%%%%%%%%%%%%%%%%%%%%%%%%%%%%%%%%%
\subsection{Spectral functions and self energies} 
For any physical system one would like to have a {\em causal\/}
description: Physical particles e.g.  may exert a measurable influence only 
after their emission. In the framework of 
quantum field theory this means that
one would like to use causal Green functions or propagators
in the theoretical description. The requirement of causality however
touches two aspects of field theory. It relates
the boundary condition in time that a propagator fulfills to
the average occupation number of the state that is propagated.

For a vacuum state, this leads to the well-known Feynman
boundary conditions, which in terms of the free propagator
in momentum state translate into the simple $+{\mathrm i}\epsilon$-
description in the denominator.

At nonzero temperature, the average occupation number of a 
state is given by a thermal equilibrium distribution function
(Bose-Einstein or Fermi-Dirac). Hence, the temporal boundary
conditions for the propagation of particles at finite temperature 
are more complicated than in a vacuum state, they are called the 
Kubo-Martin-Schwinger (KMS) condition \cite{KMS}. 

This KMS condition leads to a causal propagator with a complicated
analytical structure. It is therefore safer for thermal systems to deviate
from the description in terms of causal propagators. Rather one 
uses only retarded and advanced propagators, whose temporal boundary 
conditions do not depend on the occupation number of states. 
It is well known, how to express a finite temperature perturbation
theory in terms of retarded and advanced propagators
(see refs. \cite{KS85c,h94rep} for an extensive discussion). 

Mathematically, the retarded and advanced
propagators are analytical functions of their energy parameter 
in the upper or lower complex half-plane.
Analytical functions however obey the Kramers-Kronig relation,
and this implies that the retarded propagator of an interacting 
field theory is known completely
if only its imaginary part (or spectral function) ${\cal A}$ is known
along the real axis. Hence, for the retarded quark propagator in our
system we write, for arbitrary complex energy $E$ 
\begin{equation}\label{rapf}
S^{R,A}(E,\bbox{p})  = 
  \int\limits_{-\infty}^{\infty}\!\!dE^\prime\;
  {\cal A}_q(E^\prime,\bbox{p})\;
   \frac{1}{E-E^\prime\pm{\mathrm i}\epsilon}
\;.\end{equation}
For free particles the spectral function
is  proportional to a $\delta$-function,  
\begin{equation}\label{afree}
{\cal A}_q^{\mbox{\small free}}=( E\gamma^0 - \bbox{p}\bbox{\gamma} + m_q)\,
 \mbox{sign}(E)\,\delta(E^2-\bbox{p}^2 - m^2_q)
\;,\end{equation}
which kinematically limits the asymptotic states to be on-shell. 

However, as we have argued above, such asymptotically stable states
are not present in a finite temperature system: Each 
particle is subject to collisions which will add a statistical (thermal)
uncertainty to its energy as function of time
(thermal scattering, or Brownian motion). This indicates
that the limit of a $\delta$-like spectral function cannot be used
in interacting thermal systems -- which has been proven rigorously
in the Narnhofer-Thirring theorem \cite{NRT83}.

One may attribute this to a fundamental property of 
{\em temperature\/}: A thermal particle distribution function has a special
rest frame, hence corresponds to a breaking of the Lorentz invariance.
It is well-known that a state of broken symmetry 
requires to chose adequate basis functions
for a quantization. In case of the finite temperature breakdown
of Lorentz invariance, the basis functions turn out to be 
quantum fields without a mass shell \cite{BS75,L88}. In other words,
the field theoretically correct way to treat a finite
temperature system is in terms of {\em continuous\/} spectral functions.
 
How to put these two aspects together, i.e., the transformation to 
retarded/advanced propagation as well as the perturbative expansion in terms of
generalized free fields with continuous mass spectrum, is discussed in ref. 
\cite{h94rep}. For the purpose of the present paper, 
it is sufficient to choose a parametrization for such a spectral function.
%%%%%%%%%%%%%%%%%%%%%%%%%%%%%%%%%%%%%%%%%%%%%%%%%%%%%%%%%%%%%%%%%%%
%   SUbsection II a
%%%%%%%%%%%%%%%%%%%%%%%%%%%%%%%%%%%%%%%%%%%%%%%%%%%%%%%%%%%%%%%%%%%
\subsection{Parametrization of the quark spectral function}
For this parametrization we take as a guideline the idea to be
not too far from the quasi-particle picture, i.e.,   
we make an ansatz for the inverse retarded quark propagator 
\begin{equation}
\label{lop}
p_\mu\gamma^\mu - m_q^0 - \Sigma^R(p) \approx 
 (p_0\pm {\mathrm i}\gamma_q)\gamma_0 - \bbox{p}\bbox{\gamma} - m_q
\end{equation}
with a given self energy function $\Sigma^R(p)$ in the
vicinity of $p_0=\pm\sqrt{\bbox{p}^2 + m^2_q}$ and 
$\left|\bbox{p}\right|\ll m_q$. This implies, that we assume
the whole model to be dominated by its infrared sector, see
the remark at the end of this subsection.
This ansatz translates into a spectral function as
\begin{equation}
\label{eqaq} 
{\cal A}_q(E,\bbox{p}) = \frac{\gamma_q}{\pi} 
\frac{\gamma_0\left(E^2 + \Omega_q(\bbox{p})^2\right) -
      2 E \bbox{\gamma}\bbox{p} + 2 E m_q}{
  \left(E^2 - \Omega_q(\bbox{p})^2\right)^2 + 4 E^2 \gamma_q^2}
\;.\end{equation}
Here, $\gamma_\mu = (\gamma_0,\bbox{\gamma})$ is the four-vector of
Dirac matrices and
 $\Omega_q(\bbox{p})^2 = \bbox{p}^2 + m^2_q + \gamma_q^2$.
$m_q$ is the dynamical mass of the quark, and its spectral width parameter
we label $\gamma_q$. Note however, that the half-maximum width of the
spectral function peak is $2\gamma_q$.

One may regard this spectral function as the generalization of the
standard energy-momentum relation of eq. (\ref{afree}) to a broader 
distribution for thermally scattered particles, in this particular 
case represented by a double Lorentzian. Note also, that 
this parametrization differs from a quasi-particle approximation
only by one parameter $\gamma_q$, and in the limit 
$\gamma_q\rightarrow 0$ one recovers the free spectral function
(\ref{afree}).

Moreover, it may be shown explicitly, that this spectral function has
a four-dimensional Fourier transform that vanishes for spacelike
coordinate arguments \cite{specs}. 
Since the Fourier transform of the spectral
function is nothing but the expectation value of the anti-commutator
function of two quark fields, this is an important aspect: It guarantees,
that there cannot be any propagation of interactions faster than light.
Let us note, that with a general momentum dependence of the
spectral width parameter, this requirement may be violated.

For the self energy function we use expressions obtained in a 
skeleton expansion of the full Green function, i.e., we employ 
Feynman diagrams for this self energy which are again functionals 
of the spectral function we wish to determine.  

\begin{figure}[t]
\setlength{\unitlength}{1mm}
\begin{picture}(150,45)
\put(20,20){$p=(p_0,\vec{p})$}
\put(90,20){$k=(k_0,\vec{k})$}
\end{picture}\\
%%
%% dvips
\includegraphics{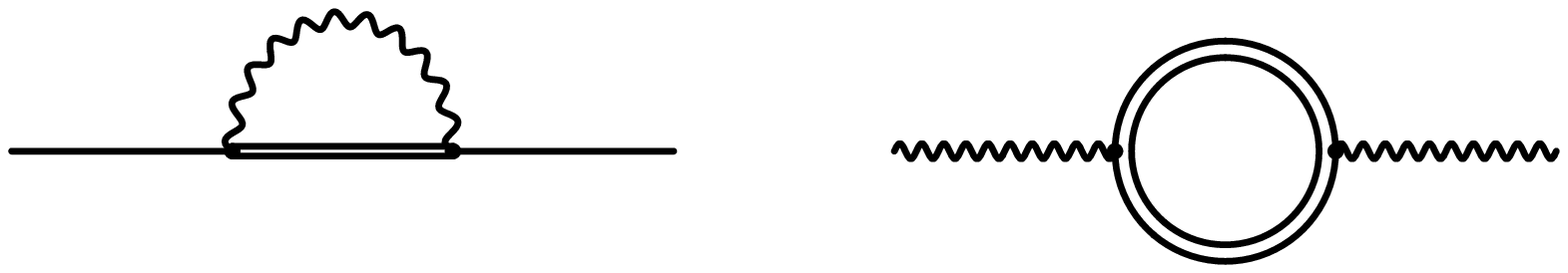}
\caption{Self energy diagrams for the photon production rate.}
\label{qedfeyn}
Left: Fock diagram for the quark self energy contribution,\\
right: photon polarization diagram including full fermion propagators on 
\underline{both} lines.
\\
\hrule
\end{figure}

In such an expansion, the one-loop (Fock) diagram,
depicted in fig.~\ref{qedfeyn}, is the lowest order term 
with a non-vanishing imaginary part. In the following, we restrict ourselves 
to this lowest order. We consider a model where quarks are coupled to 
different types of bosons, to be specified later. The calculation of 
the Fock self energy with full propagators is straightforward \cite{h94rep} 
and gives for the imaginary part
\begin{eqnarray}
\label{eqimsi}
&&{\mathrm Im} \Sigma^R(p_0,\bbox{p}) =\\
\nonumber
&&\;\;\;\;-\pi\,\int\!\!\frac{d^3\bbox{k}}{(2\pi)^3}\,
   \int\limits_{-\infty}^\infty\!\!dE\;\Gamma_{\mu}\,
    {\cal A}_q(E,\bbox{k})\,\Gamma_{\nu} \;
{\cal A}^{\mu\nu}_B(E-p_0,\bbox{k}-\bbox{p})\,\;
    \left(n_q(E)+n_B(E-p_0)\right)
\;.\end{eqnarray}
Here, ${\cal A}_B$ is the boson spectral function, $\Gamma_\mu$ and 
$\Gamma_\nu$ are the interaction matrices at the vertices, 
and $n_B$ ($n_q$) is the standard thermal equilibrium 
Bose (Fermi) distribution functions at temperature $T$, 
\begin{equation}\label{nbdf}
n_{B,q}(E) = \frac{1}{\displaystyle{\mathrm e}^{\beta E}\mp 1}
\; .
\end{equation}
The real part of this self energy function is determined by
a dispersion integral, similar to (\ref{rapf}) for the propagator.
Note, that the divergence of this integral either requires
renormalization or a regularization procedure.

Having specified the self-consistency criterion for the
quark propagator, we may now ask for its validity. In particular,
one may suspect that representing the complicated quark spectral 
function over the whole range of energies and momenta by only two parameters
is an oversimplification. However, we find on the 
contrary that for the self-consistent Fock approximation
with {\em massless\/} vector bosons the fermion spectral width 
is dominated by a constant term \cite{hsw94}. 
Hence, at low temperatures  our ansatz for the spectral
function is consistent for quark momenta $|\bbox{p}|<m_q$.

For higher temperatures the quark mass is small, 
whereas the quark momenta are typically of the 
order of the temperature. However, as has been shown in ref. \cite{PPS93},
the damping rate is dominated by the minimal distance in the complex
energy plane between the origin and the 
spectral function pole (note, that physical {\em propagators\/} do not
have poles in the complex plane). This minimal distance is
again given by the width, see eq.~(\ref{ppse}).

For the loop integrals in self energy functions the limitation
to small quark momenta is in principle violated. 
However, equilibrium distribution functions
effectively provide a cutoff at momenta $|\bbox{p}|\simeq T$.  
We therefore find, and have confirmed this by extensive numerical
computations, that the ansatz of a momentum independent spectral width
parameter is very well justified for temperatures
$T\stackrel{<}{\sim} \sqrt{m^2_q +\gamma_q^2}$ -- a relation,
which is satisfied in our approach. Only in the limit of
asymptotic freedom, where the coupling parameters indeed become small, this 
approximation will possibly fail.
%%%%%%%%%%%%%%%%%%%%%%%%%%%%%%%%%%%%%%%%%%%%%%%%%%%%%%%%%%%%%%%
%%%%%%%%%%%%%%%%%%%%%%%%%%%%%%%%%%%%%%%%%%%%%%%%%%%%%%%%%%%%%%%
% Subsection II b
%%%%%%%%%%%%%%%%%%%%%%%%%%%%%%%%%%%%%%%%%%%%%%%%%%%%%%%%%%%%%%%
\subsection{Gauge invariant photon production rates}
The width calculated from the quark self energy diagram now enters the photon 
polarization $\Pi$ at finite temperature, see fig.~\ref{qedfeyn}.
The imaginary part of the retarded one-loop polarization function 
$\Pi^R$ is \cite{h94rep}
\begin{eqnarray}
\nonumber
{\mathrm Im} \Pi^R_{\mu\nu}(k_0,\bbox{k}) &  = &
 -\pi\, e^2_q\,\int\!\!\frac{d^3\bbox{p}}{(2\pi)^3}\,
   \int\limits_{-\infty}^\infty\!\!dE \\ 
&&     \mbox{Tr}\left[\gamma_{\mu} {\cal A}_q(E+k_0,\bbox{p}+\bbox{k})
    \gamma_{\nu} {\cal A}_q(E,\bbox{p}) \right]\,\left(n_q(E)-n_q(E+k_0)\right)
\;,\label{eqimpi}
\end{eqnarray}
where $e_q$ is the electric charge of the quark. The photon production
rate for the hot plasma is proportional to this imaginary part,
summed over the different physical photon polarization directions.

We now address the question of gauge invariance of the rate calculated in 
this manner. The photon production rate is gauge invariant if the current 
which produces the photons is conserved. For the current conserving 
polarization tensor, which we denote by $\widetilde{\Pi}$, this implies 
transversality, $k^\mu \widetilde{\Pi}_{\mu\nu}(k) = 0$. 

The polarization tensor $\Pi$ as calculated from eq.~(\ref{eqimpi}), 
which is connected to the current-current correlator $\Pi_{\mu\nu}(x,y) 
\propto \left<\overline{\psi}_x \gamma_\mu\psi_x\cdot\overline{\psi}_y 
 \gamma_\nu\psi_y \right>$, violates this requirement, and may not in 
general be used for a calculation of the photon production rate. 

This can be traced back to the fact, that the naive current 
$\overline{\psi} \gamma_\mu\psi$ is not conserved. 
Of course, a theory with a nontrivial spectral 
function also has a conserved (electromagnetic) current -- but 
this differs from the naive expression \cite{hb95}. 

Let us briefly discuss the nature of this difference, starting
from the lagrangian of a generalized free field which gives rise to
a propagator with certain self energy insertion. A detailed discussion
is carried out in ref. \cite{hb95}. For a one-component
fermion field, this would be 
\begin{equation}
{\cal L}[\psi] = \overline{\psi}(x)\left({\mathrm i}\partial_\mu\gamma^\mu
 - m_0 \right)\psi(x) - \int\!\!d^4y\,\overline{\psi}(x)\, \Sigma(x,y)
\,\psi(y)
\;.\end{equation}
Performing a local phase transformation of this field then allows
to find a conserved current 
\begin{equation}
j^\mu(x) = \overline{\psi}(x)\gamma^\mu\psi(x) 
 -{\mathrm i}\int\!\!d^4y\,d^4z\; \overline{\psi}(z)\, \Lambda^\mu(z,y;x)
\,\psi(y)
\;.\end{equation}
The function $\Lambda^\mu$ is a vertex correction function. Current 
conservation now is equivalent to the
fulfillment of the Ward-Takahashi identity, which for the Fourier
transformed quantities reads 
\begin{equation}
(p - q)_\mu \Lambda^\mu(p,q) = \Sigma(p) - \Sigma(q)
\;.\end{equation}
In this equation, $p-q = k$ is the photon four momentum.
Without loss of generality we may fix the photon 3-momentum to be 
the vector $(0,0,k)$. It is then obvious, that the Ward identity
involves only the components $\Lambda^0$ and $\Lambda^3$, it does not 
restrict the transverse parts $\Lambda^1$ and $\Lambda^2$ of the vertex 
correction function. In short words, the logical steps are: \\
$\Lambda^0, \Lambda^3$ nontrivial $\Rightarrow$ 
transversality of the polarization tensor $\widetilde{\Pi}$ $\Rightarrow$ 
current conservation $\Rightarrow$ gauge invariance of
the photon production rate.

Naturally this does not imply, that vertex corrections --
if calculated diagrammatically -- have only $\Lambda^0, \Lambda^3$ 
components: It merely tells us, what is {\em sufficient\/} to ensure
gauge invariance of the photon production rate. Specifically for
the spectral function we have postulated, only $\Lambda^0$ 
is necessary to acquire a conserved current.
 
Correspondingly only the components 
$\widetilde{\Pi}^{0\nu}=\widetilde{\Pi}^{\nu 0}$ of the current 
conserving polarization tensor are different from the one-loop expression 
(\ref{eqimpi}). This tensor is the autocorrelation function 
$\widetilde{\Pi}_{\mu\nu} \propto \left<j_\mu\;j_\nu \right>$ 
of the conserved current. It is crucial to realize that the 
space-like components are not modified, $\widetilde{\Pi}^{ij} = \Pi^{ij}$.

In the next step we use this fact together with the condition
of on-shell photons, $k_0=|\bbox{k}|$. Current conservation
implies that $k^2_0\,\widetilde{\Pi}_{00}=|\bbox{k}|^2\,
\widetilde{\Pi}_{33}$, i.e., $\widetilde{\Pi}_{00}= \widetilde{\Pi}_{33}$
for on-shell photons. Hence, these two components cancel 
in the sum over polarizations:
\begin{equation}
 \widetilde{\Pi}_{\mu}^{\mu} = \widetilde{\Pi}^{00} - \widetilde{\Pi}^{ii}
 = \widetilde{\Pi}^{00} - \Pi^{ii} = -(\Pi^{11}+\Pi^{22})
\;.\end{equation}
Let us note, that this chain of arguments is rigorous: One may debate, 
whether vertex corrections are necessary out of phenomenological 
reasons -- but they are {\em not\/} necessary in order to achieve 
a gauge invariant result for the on-shell photon 
production rate. Of course, this conjecture has a drawback: 
For the more general case of a momentum dependent spectral function the 
argument above does not hold any more. Also the calculation 
of off-shell photon production, such as required for dilepton production 
rates, necessitates the calculation of vertex corrections. 

To summarize this discussion: On the level of our approximate spectral
function, and with the fully causal propagators following from this
spectral function, the gauge invariant
photon emission rate out of the hot plasma is
\begin{equation}
 R(E_{\gamma},T) = E_{\gamma} \frac{ dN_{\gamma}}{d^3\bbox{p}} = 
 2\frac{n_B(E_\gamma,T)}{8 \pi^3}\,
 \mbox{Im}\left(\Pi^R_{11}+\Pi^R_{22}\right)
=\frac{{\mathrm i}}{8 \pi^3}\,\left(\Pi^<_{11} + \Pi^<_{22}\right)
\;,\label{prate}
\end{equation}
where $\Pi^<$ stands for the off-diagonal component in the
standard notation of real-time thermal field theory \cite{h94rep}.
 
The polarization tensor $\Pi(x,y)$ is the correlator of electromagnetic 
currents at different space-time points. Hence, interference effects between
photons emitted from different points in space and time, 
as far as they affect the single photon rate, are automatically 
taken into account. This also includes the Landau-Pomeranchuk-Migdal (LPM) 
effect: Multiple thermal scattering of the slowly moving quarks leads to the 
interference of sequentially emitted soft photons, thereby reducing the soft 
photon rate. The equivalence of the semi-classical LPM description with the 
field theoretical formulation used here has been proven in ref. \cite{kv95}. 
We will show that it is exactly this interference effect which gives rise to 
our primary result. 
%%%%%%%%%%%%%%%%%%%%%%%%%%%%%%%%%%%%%%%%%%%%%%%%%%%%%%%%%%%%%%%%%%%%%%%%%%
% Subsection II c
%%%%%%%%%%%%%%%%%%%%%%%%%%%%%%%%%%%%%%%%%%%%%%%%%%%%%%%%%%%%%%%%%%%%%%%%%%
\subsection{QED plasma and astrophysical consequences}
%%%%%%%%%%%%%%%%%%%%%%%%%%%%%%%%%%%%%%%%%%%%%%%%%%%%%%%%%%%%%%%
\begin{figure}[t]
\vspace*{110mm}
%%
%% dvips
\includegraphics{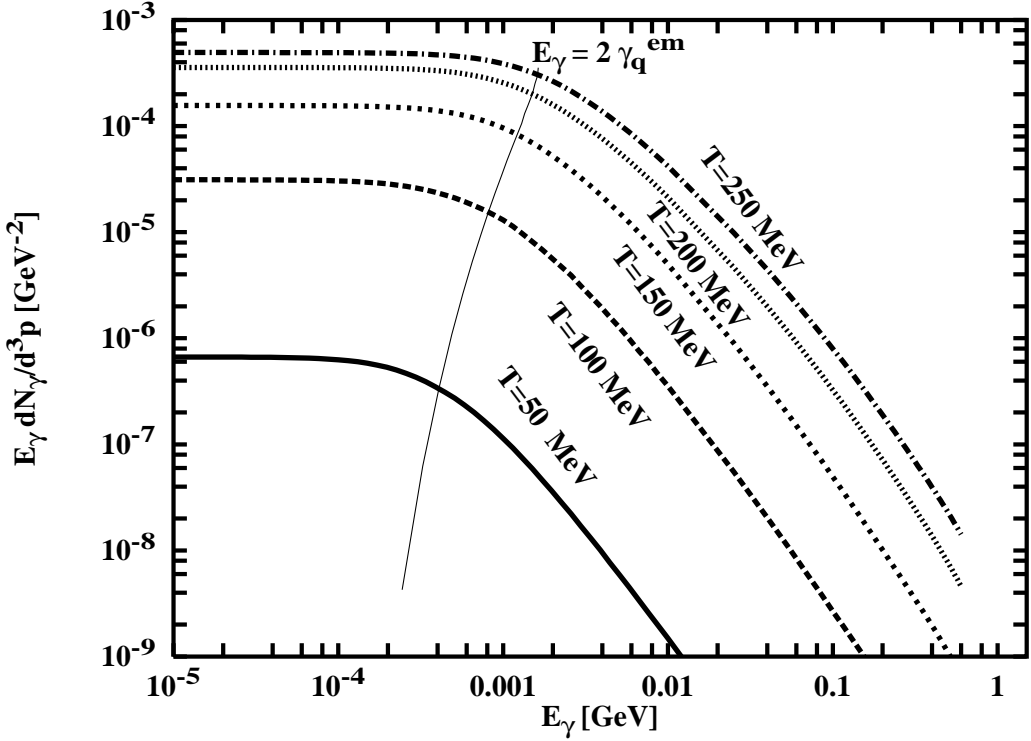}
\caption{Photon production rate $R_{\gamma}$ from an electromagnetically 
interacting particle of 300 MeV mass in a plasma as a function of the 
photon energy $E_{\gamma}$ for different temperatures $T$.}
\label{relm}
\hrule
\end{figure} 
%%%%%%%%%%%%%%%%%%%%%%%%%%%%%%%%%%%%%%%%%%%%%%%%%%%%%%%%%%%%%%%
In this subsection, we illustrate the physics with the example of a fermion
(quark, or ``heavy electron'') 
interacting electromagnetically in a plasma of temperature $T$. 
This can be handled to a large extent analytically and thus allows for a 
clearer understanding of the mechanism we are treating. 

For the purpose of this example, we keep in the following the mass of the 
fermion fixed at an (arbitrary) value of 300 MeV, and solve the 
self consistent equation of the in medium quark propagator, eq.(\ref{lop}), 
for the width only, using using real and imaginary part 
of the Fock self energy (\ref{eqimsi}) for massless free photons. 
The cutoff for the determination of the real part of the 
self energy function is chosen as $\Lambda+\sqrt{\Lambda^2+m^2_q}$, 
with $\Lambda$ = 650 MeV (see the discussion in section III.A). 

Although this width $\gamma^{\mbox{\small em}}$ is, in principle,
a non-analytical function of the temperature \cite{hsw94}, the smallness of 
the electromagnetic coupling constant allows to approximate 
it very well by the lowest order result 
\begin{equation} \label{gamelm}
 \gamma_q^{\mbox{\small em}}(T) \approx \frac{5}{9} \alpha T \cdot 
 [1 - \frac{{\mathrm Re}\Sigma^{V}}{m_q}] \approx \frac{5}{9} \alpha T 
 \cdot 
 [1 - \frac{10}{9} \frac{\alpha}{\pi} \frac{\Lambda T}{m_q^2} ]
\sim \frac{5}{9}\,\alpha \, T
\;,\end{equation}
where ${\mathrm Re}\Sigma^{V}$ is the Lorentz vector ($\propto \gamma^0$)
component of the fermion self energy function, and the factor 
$5/9$ is due to the (u,d)- family averaging of electric charge.
The purely electrodynamic quark spectral width as function of 
temperature is plotted in fig.~\ref{qwidth}, together
with the other contributions to be discussed later.

The photon production rate we obtain from eq. (\ref{prate}) 
with this width is shown in fig.~\ref{relm} for various typical values of the 
temperature. For small photon energies, i.e.~very soft photons, we find a 
saturation of the rate below values of 
$E_{\gamma} = 2\gamma_q^{\mbox{\small em}} \approx 2\cdot5/9\cdot \alpha T$. 
The factor 2 arises,
because the half-maximum width of the Lorentzian spectral function peak
is $2\gamma_q$ in our choice of parameters.

The physical interpretation of this effect is obvious: The emission of
low-energy photons requires unperturbed propagation of the emitter
over the wavelength of the photon. Along its path however the quark is 
subject to thermal perturbations -- and this hinders the photon emission 
for for $E_\gamma < 2 \gamma_q$. Our result agrees 
qualitatively with the conjecture of Weldon \cite{wel94}. Moreover,
we could clarify the dominant suppression scale to be 
twice the spectral width parameter of the emitting particle, or equal
to the half-maximum width of the spectral function peak.
In the spirit of the last remark in section II.B, we may
state that this is an interference effect between photons from
different points in space and time.

The rate for high energy photons falls off with photon energy $E_{\gamma}$ as 
$e^{-E_{\gamma}/T}$ from the Boltzmann factor. For a particle mass 
$m_q \gg \gamma_q$ this result coincides with previous calculations of 
eq.(\ref{eqbrpi}) with equivalent parameters. However, in contrast to this
calculation, our result does not employ singular behavior in 
the limit of $m_q \to 0$. This illustrates nicely 
how the finite thermal particle width regulates the infrared behavior. 

The photon production rate may be approximated as
\begin{equation}\label{cuo}
  R^{\gamma}_{\mbox{\tiny fit}} = \frac{ 4 \gamma_q }{E^2_{\gamma} + 
  4 \gamma^2_q}\, {\mathrm e}^{-E_{\gamma}/T}\,z[T]\;\;\; , \;\;z[T]
  \propto \left\{ {\array{lll} T^2 & \mbox{for} & E_{\gamma} \ll 2 \gamma_q\\
                               T   & \mbox{for} & E_{\gamma} \gg 2 \gamma_q
                   \endarray} \right.
\;.\end{equation}
For all temperatures, the limit $E_{\gamma} \to \infty$ is determined 
by the Boltzmann factor $e^{-E_{\gamma}/T}$. Note, that this 
functional dependence is a fitted result after the self-consistent
calculation, hence $\gamma_q$ is not an external parameter that 
my be varied independently of $T$ and the coupling strength.
Numerical fits will be published separately.

Another remark may be appropriate as to compare this result with 
standard (semi-classical) treatments of the Landau-Pomeranchuk-Migdal
effect: Although the exact value of $\gamma_q$ may be debated,
the general form for the radiation rate we obtain conforms very well
with all the semi-classical approaches \cite{kv95,CGR93}.

Let us briefly consider the astrophysical consequences of this result. For 
this we regard an era of the universe, where the photon energy density 
dominates the matter energy density, roughly 1 MeV $\le T \le 10$ MeV. 
Using eq. (\ref{cuo}) together with 
eq. (\ref{gamelm}), we find that the photon number density in the early 
universe is roughly $n_\gamma\approx 2/\pi^2\cdot T^3$, while the photon 
energy density is given by $\rho_\gamma\approx 6/\pi^2 \cdot T^4$. These 
have to be compared to the standard Bose-Einstein values of 
$n^0_\gamma \approx 2.4/\pi^2\cdot T^3$,  
$\rho^0_\gamma \approx 6.49/\pi^2\cdot T^4$. 
This comparison implies, that the effective number of degrees of freedom, 
$g^\star$ in the total energy density 
\begin{equation}
\rho = \frac{\pi^2}{30}\, g^\star\,T^4
\end{equation}
is somewhat reduced by the coherence effects we are considering. This 
then would lead to a faster expansion of the universe during the time where 
our result applies. However, this effect is on the order of a few percent and 
therefore in the moment beyond the reach of experimental observation. 

Furthermore we may assume, that in the present spectrum of the cosmic
background radiation the photons have retained the spectral cutoff
point from the moment when the universe became transparent. If we
assume this transition to happen at a temperature of $\approx$ 1 eV, 
the effect we are proposing predicts the cosmic microwave background
to be thermal at wavelengths shorter than $\lambda_{\mbox{\small cut}}=
\pi/{\alpha T}$, but leads to a reduction of the long-wavelength
background radiation below its black-body value for larger wavelengths.
Due to the expansion factor of $\simeq$ 1000, we estimate the 
present cutoff wavelength to be approximately 
$\lambda_{\mbox{\small cut}}\approx$ 1 m, which
makes it difficult to observe this deviation.
%%%%%%%%%%%%%%%%%%%%%%%%%%%%%%%%%%%%%%%%%%%%%%%%%%%%%%%%%%%%%%%
%
%   Section III
%
%%%%%%%%%%%%%%%%%%%%%%%%%%%%%%%%%%%%%%%%%%%%%%%%%%%%%%%%%%%%%%%
\section{Photon production from a strongly interacting plasma}
We now come to the main topic of this work, the production of photons 
from a QGP. For this purpose, we distinguish two temperature regimes. In the 
low $T$ region, below and around the phase transition temperature, 
dynamical chiral symmetry breaking and its restoration at $T_c \sim 200$ MeV 
plays an important role and has to be incorporated in a realistic 
description of the quark dynamics. We do so by using the 
Nambu--Jona-Lasinio model as an effective model up to $T\sim T_c$,
and add to this model the self-consistent summation of 
quark-photon Fock diagrams.

In the high temperature limit, chiral symmetry is restored and the coupling 
is small enough for a perturbative expansion in the strong coupling constant. 
In this region, we therefore use a self-consistent determination of 
the quark width obtained in perturbative QCD \cite{sh95qcd}.

In both regimes, nontrivial spectral functions for the respective interacting 
bosons are used. 
%%%%%%%%%%%%%%%%%%%%%%%%%%%%%%%%%%%%%%%%%%%%%%%%%%%%%%%%%%%%%%%
% Subsection II a
%%%%%%%%%%%%%%%%%%%%%%%%%%%%%%%%%%%%%%%%%%%%%%%%%%%%%%%%%%%%%%%
\subsection{Nonperturbative temperature regime}
Here, we consider the 
Nambu--Jona-Lasinio model \cite{njl} in the SU(2) version on the quark 
level, see \cite{san1} for a review and the notations used in the following. 

In this effective field theory, the nonperturbative interaction between 
quark and antiquark fields at low 
momentum transfer is described by the sum of a scalar and a pseudoscalar local 
interaction, 
${\cal L}_{int} = G[(\Psi(x)\bar{\Psi}(x))^2 + 
(\Psi(x) i \gamma_5 \bbox{\tau} \bar{\Psi}(x))^2 ] $ 
where $\Psi = (u,d)$. This is understood to be 
a summation of nonperturbative gluon interactions among the quark fields. 
It models the chiral symmetry properties of QCD in the nonperturbative 
regime, which is essential to address processes on the energy scale of the 
temperature. 
%%%%%%%%%%%%%%%%%%%%%%%%%%%%%%%%%%%%%%%%%%%%%%%%%%%%%%%%%%%%%%%  
\setlength{\unitlength}{1mm}
\begin{figure}[t]
\begin{picture}(150,30)
\put(75,20){\large =}
\put(15,5){\large +}
\put(82,5){\large +}
\put(25,26){\large $(m_q, \gamma_q^{\mbox{\tiny em}})$}
\put(95,26){\large $(m_0, 0)$}
\includegraphics{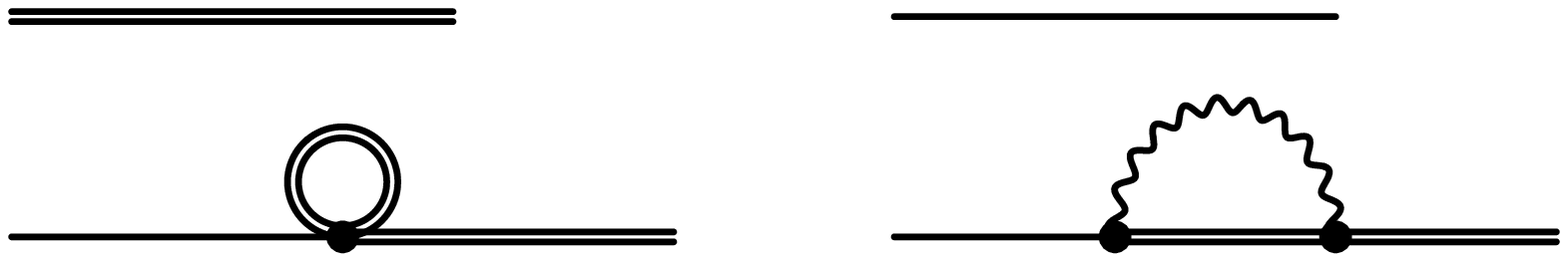}
\end{picture}
\caption{Diagrammatic equation for the quark propagator at low temperatures.}
{\small
Double line = effective quark propagator, wavy line = photon.
}
\\
\hrule
\end{figure}
%%%%%%%%%%%%%%%%%%%%%%%%%%%%%%%%%%%%%%%%%%%%%%%%%%%%%%%%%%%%%%%

This effective field theory models the chiral symmetry properties of QCD in 
the nonperturbative regime by a quartic self-interaction of quarks. At small 
temperature, the dominant contribution to the quark self energy is the 
tadpole (Hartree) term, which is expressed in terms of the spectral function
as \cite{h94rep}
\begin{equation}
 \Sigma^H = - 2 G N_C N_f\, \int\!\!\frac{d^3\bbox{p}}{(2\pi)^3}\,
  \int\limits_{-\Lambda_q}^{\Lambda_q}\!\!dE\, 
  \mbox{Tr}\left[{\cal A}_q(E,\bbox{p})\right] \, n_q(E) \; .
 \label{eqmh}
\end{equation}
Like any nonrenormalizable model, the NJL requires a momentum cutoff 
$\Lambda$, which can be motivated as a crude incorporation of asymptotic 
freedom at large $Q^2$. For the present generalization, this cutoff
is shifted to the energy integration,
\begin{equation}\label{lq}
\Lambda_q = \sqrt{\Lambda^2 + m^2_q(T)}
\;.\end{equation}
Usually, the temperature dependent quark mass $m_q(T)$ is the solution 
of the gap equation $m_q = m_0 + \Sigma^H(m_q)$. 
With appropriate parameters, this describes the 
scenario of spontaneous chiral symmetry breaking, i.e., 
the transition from a current quark mass $m_0 \approx 5$ MeV 
to the constituent quark mass $m_q \approx 1/3\; \times$ 
the nucleon mass, and its restoration at a transition temperature $T_c$. 
The only parameters were chosen as
$m_0 = 5$ MeV, $\Lambda = 0.65$ GeV and $G = 5.1\; \mbox{GeV}^{-2}$, 
and result in $T_c = 193$ MeV  and 
a vacuum mass of $m_{\pi} = 140$ MeV for the pion and $m_q$ = 332 MeV
for the constituent quark (135 MeV resp. 331 MeV without photons). 
%%%%%%%%%%%%%%%%%%%%%%%%%%%%%%%%%%%%%%%%%%%%%%%%%%%%%%%%%%%%%%%
\begin{figure}[t]
\begin{picture}(150,40)
\put(10,10){\large $\gamma_q^{\mbox {\small NJL}}$}
\put(20,10){\large $=$}
\put(27,10){\large $\gamma_q^{\mbox{\small em}}$}
\put(37,10){\large +~~Im}
\put(52,10){\Large $\biggl[$}
\put(82,10){\large +}
\put(119,10){\Large $\biggr]$}
\put(72,25){\large $\sigma$}
\put(105,25){\large $\pi$}
\includegraphics{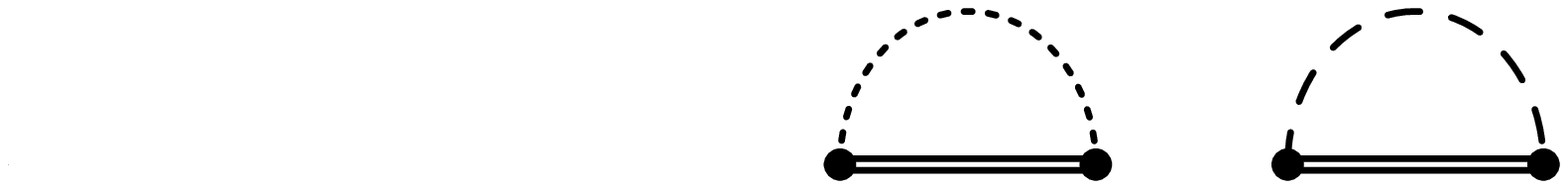}
\end{picture}
\caption{Contributions to the width $\gamma_q$ of a quark.}
\label{scwid}
{\small
Short dashed line = scalar meson, long dashed line = pion.
}\\ \hrule
\end{figure}
%%%%%%%%%%%%%%%%%%%%%%%%%%%%%%%%%%%%%%%%%%%%%%%%%%%%%%%%%%%%%%%  

The Fock self energy is the next-to-leading 
order contribution in a $1/N_c$ expansion \cite{ncp}. 
We consider quarks with four-momentum $(p_0,\bbox{p})=(m_q,0)$, hence 
we can decompose the {\em retarded\/} Fock contribution to the self 
energy in a (complex) scalar and a vector part as 
$\Sigma^{\mbox{\footnotesize Fock}} = \Sigma^S + \gamma_0 \Sigma^V$. 

The photon Fock contribution is added to the Hartree self energy, and instead 
of the gap equation we solve eq. (\ref{lop}) for the mass and width 
of the effective quark field. Split into real and imaginary part
this reads
\begin{eqnarray} \nonumber
m_q \gamma_q & = & -\mbox{Im}\Sigma^V(m_q,0) 
  \left( m_q - \mbox{Re}\Sigma^V(m_q,0) \right)\\
\nonumber
 & -& \mbox{Im}\Sigma^S(m_q,0) \left( m_0 + \Sigma^H +
\mbox{Re}\Sigma^S(m_q,0) \right)\\[3mm]
\nonumber
-\gamma_q^2 & = & \left( m_q - \mbox{Re}\Sigma^V(m_q,0) \right)^2
- \left( m_0 +\Sigma^H +\mbox{Re}\Sigma^S(m_q,0) \right)^2\\
\label{scf}
&-&\left( (\mbox{Im}\Sigma^V(m_q,0))^2 - (\mbox{Im}\Sigma^S(m_q,0))^2\right)
\;.\end{eqnarray}
Here, $\mbox{Im}\Sigma^V$ is an even function 
of $p_0$, $\mbox{Im}\Sigma^S$ is an odd function of $p_0$.
For $\gamma_q\rightarrow 0$, the above equations reduce to the
standard gap equation of the NJL model.

The NJL model furthermore describes two bound states which 
are regarded as effective, $T$-dependent pseudoscalar $\pi$-meson and 
scalar $\sigma$-meson.
 
The traditional way to determine the masses $m_B$ of these collective modes 
is to solve the equation \cite{san1}
\begin{equation}\label{trmm}
1 - 2 G\, \mbox{Re}\Pi(m_B,0) = 0
\;,\end{equation}  
with $\Pi$ the corresponding polarization function determined 
as a dispersion integral over eq. (\ref{eqimpi}). As a cutoff for
the dispersion integral one uses $\pm 2\Lambda_q$ according to
eq. (\ref{lq}).

Within our formalism we use the
following ansatz for the retarded boson propagator along the real axis: 
\begin{equation}
 \label{bopo}
  k_\mu k^\mu - m^2 - \Pi^R(k) = (k_0-(\omega_B(\bbox{k})-{\mathrm i}\gamma_B))
                             \,(k_0+(\omega_B(\bbox{k})+{\mathrm i}\gamma_B))
\end{equation}
with $\omega_B^2 = m_B^2 + \bbox{k}^2$. This translates into a spectral 
function as 
\begin{equation}\label{ab}
{\cal A}_B(E,\bbox{k})   =   \frac{1}{\pi}
\,\frac{2 E \gamma_B}{\left(E^2-\Omega_B(\bbox{k})^2\right)^2+
4 E^2 \gamma^2_B}
\;,\end{equation}
with  $\Omega^2_k  =  \omega_B(\bbox{k})^2+\gamma_B^2 $.

The parameters $m_B$ and $\gamma_B$ are determined from the equations
\begin{eqnarray}\nonumber
1 - 2 G\, \mbox{Re}\Pi(m_B,0) + \left( G\,\pi\sigma(m_B,0)\right)^2&=& 0\\
m_B\,G\, \pi \sigma(m_B,0) &=&\gamma_B
\;.\end{eqnarray} 
The mesonic Fock contributions to the quark self energy
are treated only perturbatively, i.e., their imaginary parts are used
to modify the quark width according to fig.~\ref{scwid} 
and their influence on the quark mass is neglected \cite{ncp}.

To check the consistency of this approximation,
we also performed a self consistent summation of the meson Fock diagrams,
which gives rise to a small correction of the constituent quark
mass as well as the meson masses in our extended NJL model.
However, such a summation breaks chiral symmetry
explicitly - and one may expect, that the mass shift is
canceled by other diagrams. For the purpose of the
present paper we therefore chose the perturbative treatment of
the meson Fock self energy contributions as described above.

Physically, our approach amounts to consider
photon emission processes, which are initiated by the interaction of the
quark with a {\em single\/} hot meson. The resulting quark width $\gamma_q$ 
is plotted in fig.~\ref{qwidth}. For low temperatures, we again find 
$\gamma \propto T$ for each channel. Up to a temperature of
$\approx$ 150 MeV, the quark width $\gamma_q$ is in fact dominated by
the purely electrodynamic contribution. This indicates, that 
photons should be taken into account even for strongly
interacting systems at such temperatures.

Due to the quasi--Goldstone mode of the pion, its contribution 
to the quark width remains 
negligible up to the Mott temperature $T_M$ = 212 MeV, which is defined by 
$m_{\pi}(T_M) = 2 m_q(T_M)$ as the point where the pion can 
dissociate in a $q\bar{q}$ pair. For $T > T_M$, the pionic contribution to 
the quark spectral width is actually dominant. Towards higher temperatures, 
the competing effect of an increase of the mass of the $\pi$ 
(now a resonance) again turns the width down. 

One may argue at this point, that in the NJL model quarks
are not confined. However, the above results may be translated
to other models as well: They represent nothing but 
a critical opalescence to photons in the vicinity of the chiral
phase transition. Hence we expect the drastic increase 
of the effective $\gamma$ around $T_c$ to be quite independent
of the model used.
%%%%%%%%%%%%%%%%%%%%%%%%%%%%%%%%%%%%%%%%%%%%%%%%%%%%%%%%%%%%%%%%%%%%
% Subsection III b
%%%%%%%%%%%%%%%%%%%%%%%%%%%%%%%%%%%%%%%%%%%%%%%%%%%%%%%%%%%%%%%%%%%%
\subsection{Perturbative QCD for high temperature plasma}
In the high temperature limit, a calculation within perturbative QCD becomes 
sensible. Furthermore, non-abelian gauge invariance becomes
an imperative of the calculation. For each of the degrees of freedom,
i.e., quarks, transverse and longitudinal gluons
and ghost fields, one has to consider an effective propagator 
similar to those we have discussed before. 

We will first report on the result of
ref. \cite{sh95qcd}, where, in the same spirit
as discussed for the electromagnetic case, a high-temperature
self-consistent QCD calculation was carried out.
%%%%%%%%%%%%%%%%%%%%%%%%%%%%%%%%%%%%%%%%%%%%%%%%%%%%%%%%%%%%%%%
\begin{figure}[t]
\begin{picture}(150,120)
\put(5,95){\large$\Pi_{\mbox{\small gluon}} =$}
\put(75,95){\large +}
\put(15,55){\large +}
\put(82,55){\large +}
\put(5,20){\large$\Sigma_q =$}
\includegraphics{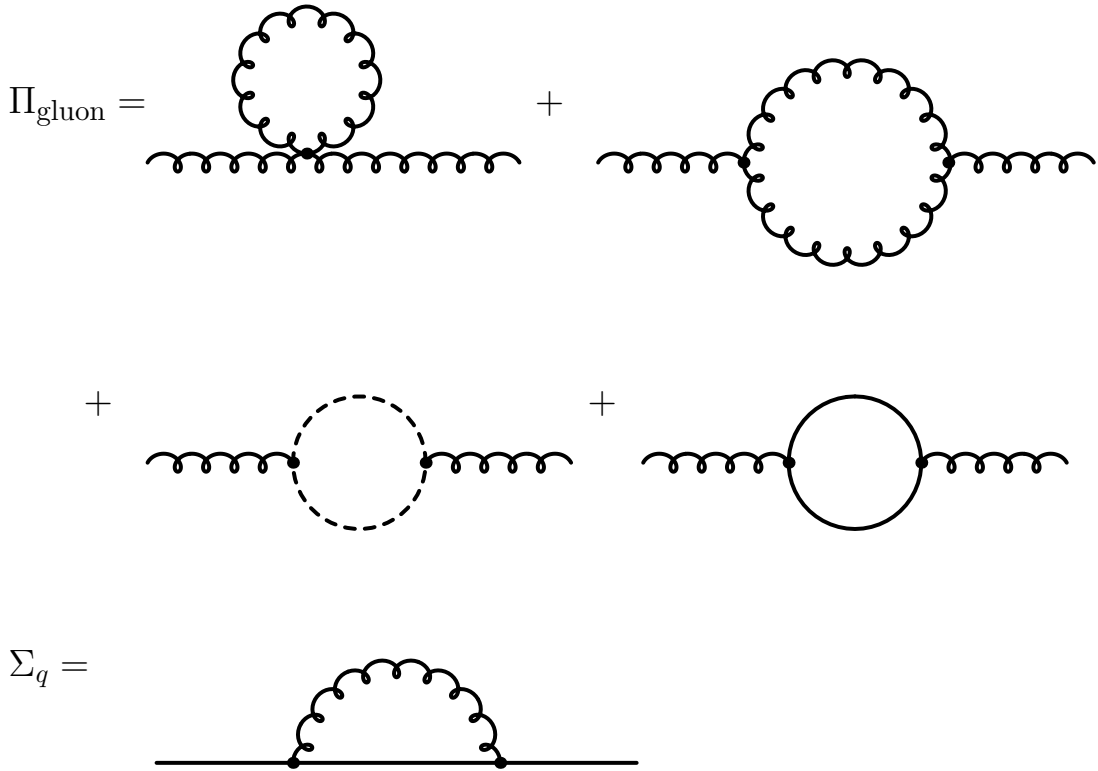}
\end{picture}
\caption{Feynman diagrams used in the calculation of the effective
quark and gluon propagator. Each line represents an
effective propagator here, dashed are the ghost fields. }
\label{qcd-feyn}
\end{figure}
%%%%%%%%%%%%%%%%%%%%%%%%%%%%%%%%%%%%%%%%%%%
\begin{figure}[t]
\vspace*{110mm}
%% dvips
\includegraphics{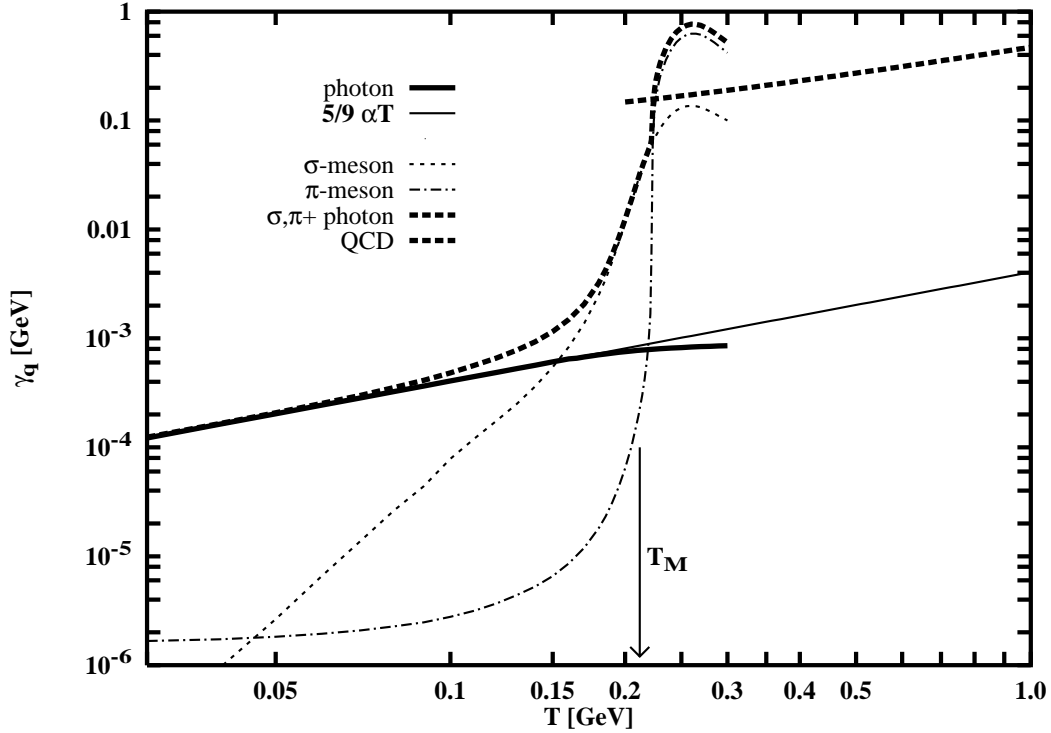}
\caption{Contributions to the width $\gamma_q$ of a quark.
Thick dashed line: total quark width, calculated with the
NJL model at low temperature and inspired by perturbative QCD at 
high temperature.}
\label{qwidth}
\end{figure} 
%%%%%%%%%%%%%%%%%%%%%%%%%%%%%%%%%%%%%%%%%%%%%%%%%%%%%%%%%%%%%%%

Various self energy diagrams have been taken into account in this work,
including -- compare this to the QED case -- a gluon Fock
diagram for the quarks as well as polarization functions for
the boson fields. The photon Fock diagram may be neglected
in this temperature regime, since its contribution
is much smaller than the gluon Fock contribution (see fig.~\ref{qwidth}).
Fig.~\ref{qcd-feyn} contains a list of the diagrams,
which were self-consistently summed in the infrared limit.
Within this calculation, the quark width is obtained as
\begin{equation} 
 \gamma_q = 0.271\, g T \;\;(\mbox{for}\; N_f =2, N_c=3) 
\;.\end{equation}
To the same order of accuracy, and in view
of the explanation in section II.B, we use the running coupling constant 
\begin{equation}
 g^2(Q^2) = 4\pi\cdot \frac{12\pi}{
\displaystyle (33-2N_f)\log\left(Q^2/\Lambda_{\mbox{\small QCD}}^2\right)} 
\end{equation}
with $N_f = 2$ for up and down flavors, only and 
$\Lambda_{\mbox{\small QCD}}$=0.2 GeV. This gives the proper match 
to the two-flavor NJL calculation towards low temperatures. Due to their 
larger current quark mass, the strange quarks will not give a large 
contribution at the temperatures we consider. We relate the mean $Q^2$ to the 
temperature by $<Q^2> \approx (3T)^2$. The resulting quark width reads 
\begin{equation} \label{shgam}
  \gamma_q = \frac{2.858\, T}{
  \displaystyle\sqrt{\log\left[\frac{15.0\cdot T}{{\mbox{GeV}}}
  \right]}} \; \approx\; 1.65\, T\;\mbox{for  $T\, \approx$ 0.2 GeV}
\;.\end{equation}
However, the question remains whether the result 
$\gamma\propto g T$ does not contradict 
the calculations in the NJL model, which yielded $\gamma\propto 
\alpha T \propto e^2 T$. On one hand, we may 
refer to eq. (33) of \cite{PPS93}, where
\begin{equation}\label{ppse}
\gamma_q\simeq \frac{3 g^2 T^2}{64 E_q}
\;.\end{equation}
Substituting for $E_q$ the minimal distance between the origin in the complex
plane and the pole of the spectral function, i.e., $E_q\simeq \gamma_q$,
we find that the self consistent $\gamma_q\simeq 0.216\, g T$.
In this framework one would therefore find, that it is the 
replacement of $m_q$ by $\gamma_q$ in the infrared screening, which
provides a $\gamma_q\propto g T$ instead of $g^2 T$.
 
On the other hand, there are other 
calculations within the hard thermal loop resummation method which
give a fermion damping rate of order $g^2 T$ \cite{BPS94}, 
although one may criticize the propagators used there because 
they violate the locality axiom of quantum field theory \cite{specs}.

There, the quark width is obtained as the solution of an
equation of the type \cite{BPS94,PPS93,BK94}
\begin{equation}
\gamma \;\simeq\;  \frac{g^2}{2 \pi^2}\,T\,\int\limits_0^{g T}\!\frac{dk}{k}\,
\arctan\left(\frac{k}{\gamma}\right)
\; =\; \frac{g^3 T^2}{8 \pi^2 \gamma} \,\Phi\left(
{\displaystyle  -\frac{g^2 T^2}{\gamma^2},2,\frac{1}{2}}\right)
\;,\end{equation}
where $\Phi(z,s,a)$ denotes the Lerch transcendent, a 
generalization of the polylogarithm function and not
expressable as a finite series of elementary functions.
Numerical evaluation gives a value of 
\begin{equation}
\gamma_q \approx 0.82\,  T \;\mbox{for $T\, \approx$ 0.2 GeV}
\;.\end{equation}
For small $g\stackrel{<}{\sim}0.5$, one finds
\begin{equation}\label{bk94imp}
\gamma \simeq \frac{g^2}{4\pi}\, T\,\log\left(\frac{1}{g}\right)\,
  \left(1 - \log\left[\log\left( \frac{1}{g}\right)
 \frac{1}{4 \pi}\right] \left\{\log\left(
  \displaystyle\frac{e}{g}\right)\right\}^{-1} \right)
\;,\end{equation}
where $e$  denotes Eulers constant. This is a
non-analytical function around the point $g=0$, but it is
{\em not\/} of the order $g^2 \log(1/g)$ as claimed
in refs. \cite{BPS94,petit,BK94} (this logarithmic piece is also denounced
in ref. \cite{PPS93}).
Moreover, there are hints that the logarithmic correction is due to
a neglection of vacuum parts in the evaluated diagrams \cite{hsw94}.

The perturbative spectral width of the quark obtained in the region of 
the chiral phase transition temperature is also compatible with 
the high temperature NJL result, which above the Mott transition temperature
gives a spectral width of $\gamma\simeq T$.

Hence, starting from three completely different methods: {\bf a.} the NJL 
model, {\bf b.} effective QCD with generalized free fields and {\bf c.} 
hard thermal loop resummation, we 
arrive at similar results $\gamma_q \approx (0.82 \dots 1.65) T$
in the region of the chiral phase transition. Each of these curves
matches with the NJL result at temperatures of $\approx 220$ MeV, as
can be inferred from fig.~\ref{qwidth}. For reasons to be made clear later
we therefore retain the parametrization of eq. (\ref{shgam}), but emphasize
again that we do not decide at this point whether the quark
damping rate is of order $g T$ or $g^2 T$. Rather, we consider this 
only as a numerical parametrization which is supported by 
{\em all\/} available methods.  

The resulting spectral width of the quark is plotted in fig.~\ref{qwidth},
together with the low temperature result obtained in the previous 
subsection. This temperature dependent spectral width of the
quark therefore has two regions which we consider to be safely established
beyond any questions of the detailed model and technique:
\begin{equation}
\gamma_q(T) \simeq \left\{ {\array{rll}
0.004\cdot T & \mbox{for} & T\ll T_c\;\; \mbox{(electromagnetic)}\\
1.0\cdot T & \mbox{for} & T\stackrel{>}{\simeq} T_M\;\;\mbox{(strong)} 
\endarray} \right.
\;.\end{equation}
The pronounced rise of the width around the phase transition temperature, 
which is carried over to a similar rise in the photon rates, is therefore 
a model independent result of finite temperature QCD. We identify it with 
the dissociation of the mesons, dominantly $\pi\leftrightarrow q \bar{q}$, 
which is connected with the phase transition temperature. 

%%%%%%%%%%%%%%%%%%%%%%%%%%%%%%%%%%%%%%%%%%%%%%%%%%%%%%%%%%%%%%%
\begin{figure}[t]
\vspace*{100mm}
%%
%% dvips
\includegraphics{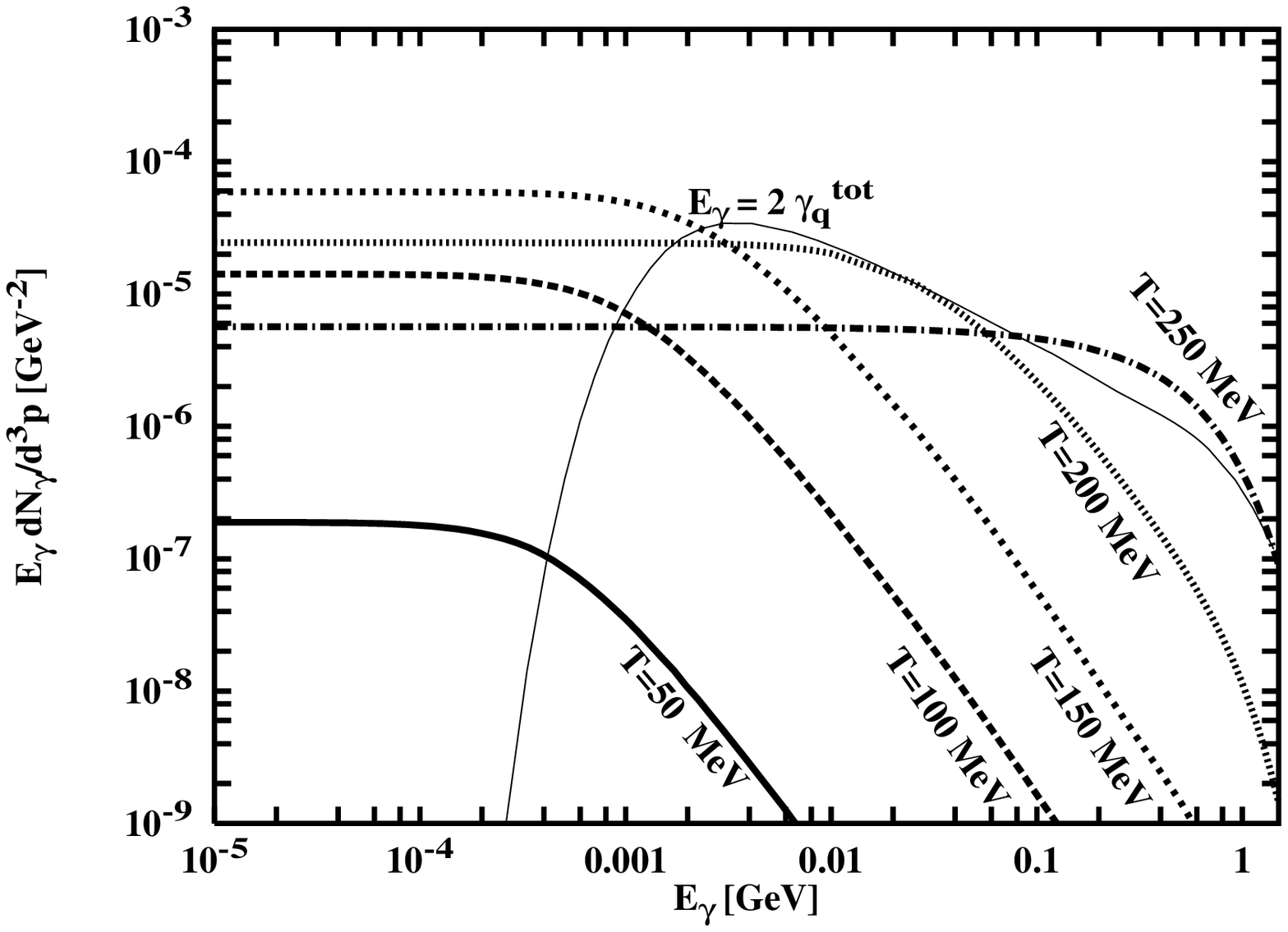} 
\caption{Photon production rate $R$ as function of
the photon energy for different temperatures.}
\label{rnjl}
{\small 
Thick lines: Temperatures 50 -- 250 MeV (dashed),
our calculation using the NJL model.\\
Thin line: Cutoff point $E_\gamma=2 \gamma_q$.
}\\ \hrule
\end{figure} 
%%%%%%%%%%%%%%%%%%%%%%%%%%%%%%%%%%%%%%%%%%%%%%%%%%%%%%%%%%%%%%%
% Subsection III c
%%%%%%%%%%%%%%%%%%%%%%%%%%%%%%%%%%%%%%%%%%%%%%%%%%%%%%%%%%%%%%%
\subsection{Results for photon production rates}
%%%%%%%%%%%%%%%%%%%%%%%%%%%%%%%%%%%%%%%%%%%%%%%%%%%%%%%%%%%%%%%
\begin{figure}[t]
\vspace*{100mm}
%%
%% dvips
\includegraphics{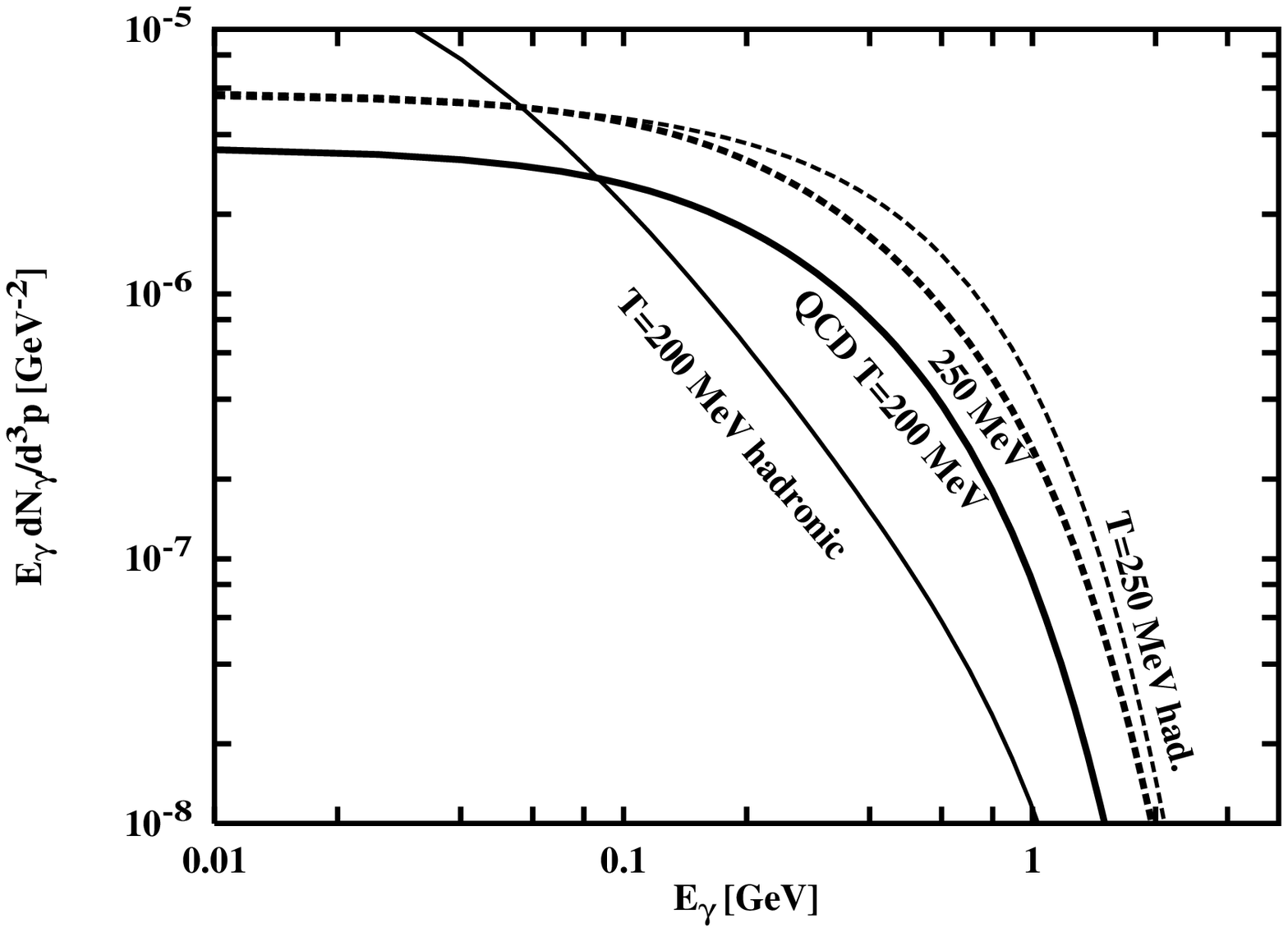} 
\caption{Photon production rate $R$ from a QGP as function of
the photon energy for different temperatures.}
\label{rtran}
{\small 
Thick lines: T=200 and 250 MeV (dashed), our calculation using 
eq. (\ref{shgam}). \\
Thin lines: T=200 and 250 MeV (dashed), our calculation using
the NJL model.
}\\ \hrule 
\end{figure}
%%%%%%%%%%%%%%%%%%%%%%%%%%%%%%%%%%%%%%%%%%%%%%%%%%%%%%%%%%%%%%%
%%%%%%%%%%%%%%%%%%%%%%%%%%%%%%%%%%%%%%%%%%%%%%%%%%%%%%%%%%%%%%%
\begin{figure}[t]
\vspace*{100mm}
%%
%% dvips
\includegraphics{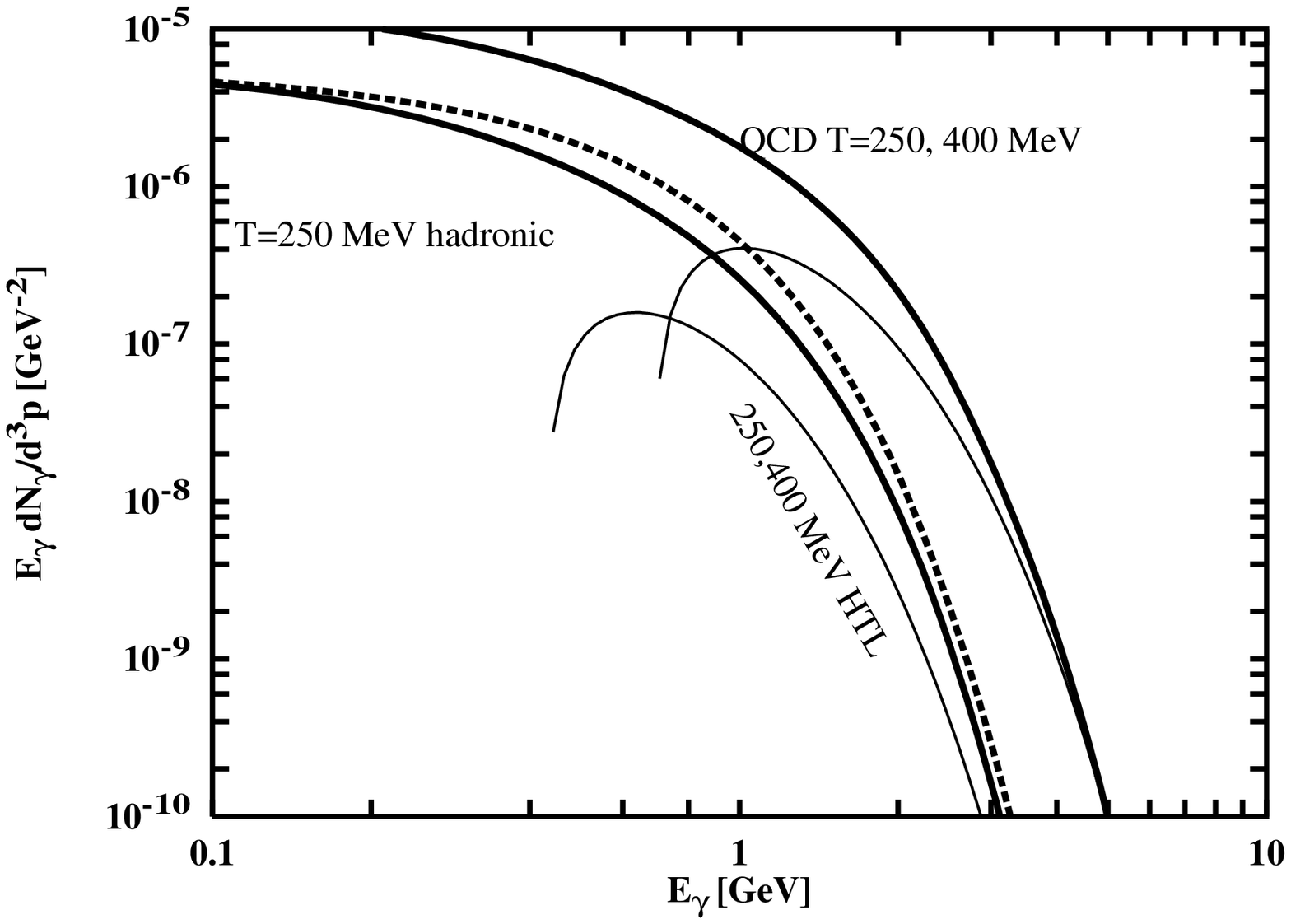} 
\caption{Photon production rate $R$ from a QGP as function of
the photon energy for two different temperatures.}
\label{rqcd}
{\small Thick solid lines: T=250 and 400 MeV, our calculation using 
 perturbative QCD, eq. (\ref{shgam}) \\
Dashed line:  T=250 MeV, our calculation using 
  the NJL model \\
Thin lines: Calculation with the method of hard thermal loops, \cite{kap1}
}\\ \hrule 
\end{figure}
%%%%%%%%%%%%%%%%%%%%%%%%%%%%%%%%%%%%%%%%%%%%%%%%%%%%%%%%%%%%%%%
\begin{figure}[t]
\vspace*{100mm}
%%
%% dvips
\includegraphics{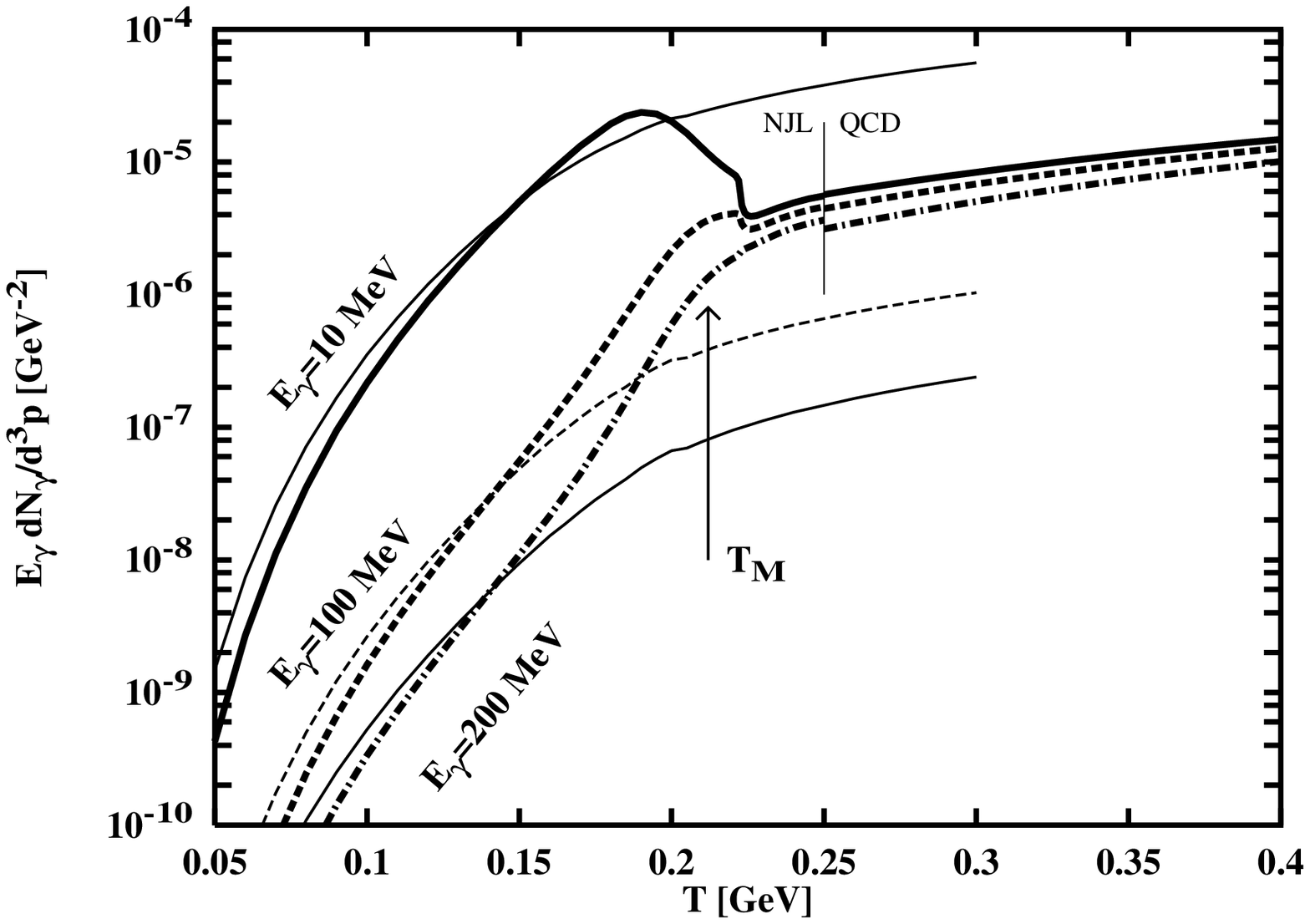} 
\caption{Photon production rate $R$ from as function of
temperature $T$.}
\label{fiqen}
{\small
Photon energy 10 MeV (solid), 100 MeV (dashed) and 200 MeV (dash-dotted).\\
Thin lines: purely electromagnetic width $\gamma_q^{\mbox{\small em}}$\\
Thick lines, curves from left: Our calculation using the NJL model\\
Thick lines, curves from right: Our calculation using 
perturbative QCD, eq. (\ref{shgam})
}\\ \hrule
\end{figure} 
%%%%%%%%%%%%%%%%%%%%%%%%%%%%%%%%%%%%%%%%%%%%%%%%%%%%%%%%%%%%%%%
First we discuss the photon emission rates obtained 
from eq.(\ref{prate}) below the chiral phase transition temperature.
They are plotted in fig.~\ref{rnjl}, and 
we find a great similarity of the rate with the results plotted in fig.~
\ref{relm} for the purely electromagnetic case.
However, at temperatures $T>T_M\approx 212$ MeV, the quark width
parameter is dominated by the mesonic contributions, which leads
to a much higher saturation scale $\gamma_q\gg \gamma_q^{\mbox{\small em}}$.
In fig.~\ref{rnjl} this rise is also documented by the 
turning of the curve $E_\gamma = 2 \gamma_q$, see the thin continuous line.

We now turn to the region of the chiral phase transition. In fig.
\ref{rtran}, we compare the NJL calculations at temperature
$T$=200 MeV (250 MeV) with the corresponding perturbative QCD
calculations. The resulting photon production rates are different
at $T$=200 MeV, i.e., very close to the chiral phase transition
temperature. The reason for this is clearly, that the perturbative
QCD calculation is no longer applicable here.

Very good agreement is reached at a temperature of $T$=250 MeV,
which is of course due to the matching values of
$\gamma_q$ at this temperature, see fig.~\ref{qwidth}.

In fig.~\ref{rqcd}, we have plotted the perturbative QCD results for
temperatures $T$=250 and 400 MeV, i.e., in a region where one would
not trust the NJL model. The figure also contains a plot
of the photon production rate obtained with the method of 
hard thermal loops, eq. (\ref{eqbrpi}). We find, that our
QCD inspired calculation at the lower temperature agrees very
well with the NJL result, while at the higher temperature and
for hard photons is reproduces the result of the hard thermal loop 
calculation. For reason of this agreement we have kept our
parametrization for the quark spectral width above $T_c$, 
eq. (\ref{shgam}).

At lower photon energies however, where according to our result
the photon radiation is cut off due to the finite mean free path
of a particle, the result of the hard thermal loop calculation
is not usable. We therefore consider our result an extension
of commonly accepted results to a wider range of energies.

The leveling off of the rates at low photon energy is due to the inclusion 
of coherence (LPM) effects. For lepton pairs, this effect can be compared to 
the results of ref. \cite{cley2}, where this effect was introduced ``by 
hand'',
whereas in our calculation it is automatically included in the formalism.

In fig.~\ref{fiqen}, we show the photon emission rate at three different
photon energies as a function of temperature. 
Comparing the electromagnetic case (thin lines)
to the model including the
quark-meson interaction, we find a surprising result: In the region of the
chiral phase transition, the low-energy photon production rate {\em drops\/} 
with increasing temperature. Eventually the radiation rate is
degenerate for all energies $E_\gamma<2\gamma_q$ (see the flat behavior
of the curves in fig.~\ref{rnjl}).  In view of eq.~(\ref{cuo}), this is 
understood as a dominance of the saturation effect over the increase 
of temperature. 
%%%%%%%%%%%%%%%%%%%%%%%%%%%%%%%%%%%%%%%%%%%%%%%%%%%%%%%%%%%%%%%%%%%%
%  Subsection III d
%%%%%%%%%%%%%%%%%%%%%%%%%%%%%%%%%%%%%%%%%%%%%%%%%%%%%%%%%%%%%%%%%%%%
\subsection{Relevance to experimental data}
We now briefly give an overview of existing or future experiments, in the 
order of increasing energy, which observe photons in reactions of hadronic 
character, and discuss the relevance of our rate calculations to them.  

Let us first consider very soft (for an experimentalist's scale) photons, 
where $E_{\gamma} \sim$ 1 -- 100 MeV. Photons in this energy region have been 
measured in several experiments \cite{helios,wa80,cergam,chl,ban}.
While some of these experiments find a complete agreement 
of the measured photon spectra with the emission from hadronic sources
and QED bremsstrahlung, an enhancement in the low $p_{\bot}$ region 
was observed in reactions such as $K^+p$ \cite{chl} and $\pi^-p$ \cite{ban}.

At the moment the discussion of experimental results is not yet 
conclusive, see ref. \cite{sgrev} for an excellent review of the data. 
However, there is one proposed explanation for such a soft photon excess 
in case it is present: It might be due to 
``thermal'' radiation of cold drops of quark--gluon plasma, which 
would hadronize only slowly and thus have a long time to radiate 
\cite{vhl,sgrev}. 

While we think it premature to draw definite conclusions from the puzzling 
experimental situation, we will comment on the proposed theoretical 
explanation of an enhanced soft radiation from a cold plasma droplet. 
Since the emitted photons 
are soft, the (LPM-) effect of interference between successive emitters 
is necessarily very strong. Using our results for the rate as shown in 
fig.~\ref{fiqen}, we therefore estimate the expected 
photon yield from such a cold plasma drop of some size $R$ at a 
temperature $T$. From uncertainty, $R \geq 1/\overline{p}$ where 
$\overline{p}$ is the mean momentum of a plasma particle, and thus cold drops 
need to be of considerable size. For a drop of volume $V$ and lifetime $\tau$ 
at a temperature $T$, the differential cross section of photon production at a 
transverse momentum $p_{\bot}$ can be expressed as 
\begin{equation} 
  \frac{d\sigma}{dp_{\bot}} = \frac{V\tau}{(0.2 fm)^4} 2 \pi p_{\bot} 
  \sigma_{AB} R(E_{\gamma},T) 
\end{equation}
where $R(E_{\gamma},T)$ is the invariant rate and $\sigma_{AB}$ is the total 
cross section of the reaction $A+B$ under study. 

As an example, let us consider the reaction $K^+ p \to \gamma X$ where the 
$p_{\bot}$ spectrum of photons was measured \cite{chl}. 
At a temperature of 50 MeV, the invariant rate of photons of say 
$E_{\gamma}$ = 10 MeV is, reading from fig.~\ref{fiqen}, 
about $10^{-9}$ mb/GeV. Taking a drop 
lifetime $\tau \sim R \sim$ 10 fm and a cross section of $\sigma_{K^+p}$ =
16 mb results in an estimate of the photon yield 
of $d\sigma/dp_{\bot} \sim 3\cdot 10^{-2}$ mb/GeV. This is to be compared 
with the measured value of $d\sigma/dp_{\bot}|_{\mbox{\small exp}} \sim 200$
mb/GeV, which is much larger than the rate estimated from a cold plasma 
drop emitting photons. 

As we noted before, previous calculations in the cold plasma
droplet picture had arrived at numbers which correspond to
the excess over the bremsstrahlung found in some experiments.
However, these calculations do {\em not\/} account for the coherence
(LPM-) effect which dramatically reduces the rate, and actually the 
normalization of the spectra obtained in these calculations was taken from 
experiment. We thus have to conclude, 
that the mechanism of ``thermal'' radiation from cold plasma droplets does
not account for an excess of soft photons over the bremsstrahlung --
independent of the question, whether such an excess is found experimentally 
or not.
 
Now consider higher energies of the photon in the range $E_{\gamma} \geq$ 100 
MeV. In recent heavy ion experiments at ultrarelativistic energies, it is 
hoped to find some hints of a phase transition the system might, possibly 
partially, go through. Apart from measuring photons, experiments also observe 
lepton pairs which suffer less from coming together with a large background. 
For both electromagnetic probes, an enhancement might hint at the new phase. 

For photons, the invariant rate as shown in figs.~\ref{rqcd} and \ref{fiqen}
gives our result for the 
QGP, and needs to be folded with the space-time evolution of the system 
such as calculated in \cite{phot}. For this purpose, the invariant rate may be 
written in terms of the photon rapidity $y$ and transverse momentum 
$p_{\bot}$ as 
\begin{equation}
E{d\sigma\over d^3\bbox{p}} = {1\over 2\pi p_t} {d\sigma\over dp_tdy} \; .
\end{equation}
Photons with a low virtuality can be converted into dileptons by use of the 
soft photon approximation, 
\begin{equation}
  R_{l^+l^-} = E_+E_-\frac{dN_{ll}}{d\bbox{p}_+^3 d\bbox{p}_-^3} \approx 
  \frac{\alpha}{2 \pi^2 M^2} E_{\gamma} \frac{dN}{d^3\bbox{k}} \; , 
\end{equation}
and improved versions thereof \cite{ruck}. This allows the use of the results 
presented in this work to the calculation of dilepton rates as well. 
Again, a space-time integration needs to be performed to compute the yield 
for a heavy ion reaction at some impact parameter, which is related to the 
measured multiplicity or total transverse energy.

This integration over the space-time evolution of the collision was performed 
by use of the photon rates calculated from hard thermal loops \cite{phot}. 
The system proceeds from an initial QGP through a mixed phase to a purely 
hadronic phase in the final state. Since the initial plasma phase is 
short-lived
and of a similar temperature than later phases under the conditions studied, 
and the QGP does not shine very much brighter than a mixed or hadronic phase 
of the same temperature, the yield of photons from the QGP is much 
smaller than that of the other phases, typically more than an order of 
magnitude. As can be seen from fig.~\ref{rqcd}, our results in the range of 
photon energies $E_{\gamma} \geq$ 100 MeV are similar to those from hard 
thermal loops which had been used in the analysis \cite{phot}. Therefore, 
the conclusion remains, that the scarce photons from 
a plasma phase are overwhelmed by those from the later stages of the reaction, 
and the same is the case for dileptons. This applies to the application of 
our result to current experiments such as Ceres, WA80/98 and Helios. In 
particular, an enhancement seen in these experiments can not be accounted for 
by a direct contribution of the plasma phase, but must be of different origin 
(which, of course, may still be related to a phase transition).

A substantial transverse flow in these collisions, for which there is some 
evidence \cite{trans1}, has the tendency to increase the apparent photon 
energy and thus to increase the photon rates from a QGP \cite{trans2}. 
However, we expect that transverse flow has a similar effect on photons 
produced from the hot hadron gas in the final stage. Therefore, this effect 
is probably of no help for the distinction of these photon sources 
in present measurements. 

This situation might change in favor of the quark--gluon plasma
when going to higher 
energies. Here the rate of electromagnetic probes originating in the plasma 
rises strongly with increasing temperature of the QGP (see fig.\ref{rqcd}), 
while the 
temperature at which the hadronic reactions occur does not change. 
Experiments which are under preparation at RHIC and LHC are planning to 
observe photons and dileptons and thus it is hoped that these experiments 
might see the QGP in sufficiently bright light in order to uniquely identify 
this phase. 
%%%%%%%%%%%%%%%%%%%%%%%%%%%%%%%%%%%%%%%%%%%%%%%%%%%%%%%%%%%%%%%%%%%%
%  Section IV
%%%%%%%%%%%%%%%%%%%%%%%%%%%%%%%%%%%%%%%%%%%%%%%%%%%%%%%%%%%%%%%%%%%%
\section{Conclusions}
In this work, we calculated gauge invariant rates for the production of 
low and high energy 
photons from a hot plasma in the framework of thermal field 
theory with generalized free fields. For illustration, we studied a 
QED plasma and discussed possible astrophysical consequences. The main 
purpose of this work, the calculation of photon rates from a 
quark--gluon plasma, was achieved over the entire range of temperatures in 
a composition of two scenarios. For high temperature, perturbative QCD has 
been used, while around the chiral phase transition region, the 
nonperturbative NJL model was employed. Both results were found to match 
smoothly. This is a very satisfactory result and might be of more general 
relevance than in this particular case. 
For high photon energies, our result are similar to those obtained 
previously in the hard thermal loops technique, which is applicable only in 
the high temperature regime.  

It was one of the main motivations of the present work to demonstrate how 
meaningful production rates may be obtained at finite temperature for soft 
photons, where the coherence (LPM) effect plays an important role. 
We emphasize that the qualitative properties of the soft photon rates, 
such as the saturation effect towards low temperatures, follow from general 
physical considerations as we discussed, and are in particular independent of 
the particular model we used. In particular, the decrease of the production 
rate of soft photons in the temperature region of the phase transition 
is a very intriguing result, which also might have observable consequences. 

We pointed out that at presently reachable energies, photon production in an 
ultrarelativistic heavy ion collision is dominated by later phases of the 
reaction rather than an initially present QGP. However, at the energies of 
currently planned experiments, the plasma temperatures might be high enough to 
allow a direct identification of this phase. Here, the precise rate for the 
production of photons of a given energy as we calculated it is a very 
important tool for the for the interpretation of the experimental results. 

From our results one may furthermore conclude, that quantum field theory 
in terms of generalized free fields with reasonable parametrizations
of spectral functions is a valuable method for the analysis of
relativistic heavy ion collisions. The strong gap between quantum field
theory as a formalism and its predictive power for
{\em many-body\/} experiments, which has persisted for some time,
is hopefully bridged by the application of {\em thermal\/} field theory. 
We are currently undertaking an effort to derive simple parametrizations
of the photon production rate that might be used as input for 
simulation codes.

\vspace*{.5cm}

{\em Acknowledgments:}\\
We would like to thank 
J.H\"ufner, J.Kapusta, J.Knoll and W.Weise for 
valuable discussions and suggestions. To D.Srivastava we owe
our thanks for pointing out the problem of soft photons,
and to H.J.Specht for clarifying this problem in a critical discussion.
%%%%%%%%%%%%%%%%%%%%%%%%%%%%%%%%%%%%%%%%%%%%%%%%%%%%%%%%%%%%%%%
%%%%%%%%%%%%%%%%%%%%%%%%%%%%%%%%%%%%%%%%%%%%%%%%%%%%%%%%%%%%%%%

\end{document}